\documentstyle[12pt]{article}
\begin{document}
\thispagestyle{empty}
\begin{center}
\LARGE \tt \bf{Torsion strings inside Static Black holes in Teleparallel Gravity}
\end{center}
\vspace{1cm}
\begin{center}
{\large L.C.Garcia de Andrade }{\footnote{Departamento de Fisica Teorica-IF-UERJ-e-mail:garcia@dft.uerj.br}}\\
\end{center}
\vspace{0.5cm}
\begin{abstract}
Cosmic strings inside Schwarzschild black holes in teleparallel gravity are considered.Torsion flux inside the black hole is compute like a torsion vortex on a superfluid.Since some components of torsion are singular on Schwarzschild horizon and others remain finite we compute a torsion invariant to decide whether the torsion is singular and where torsion singularities located.It is found out that as in the curvature case of Einstein's black hole the event horizon is not a mere coordinate singularity for torsion although a true torsion singularity is found at the center of the teleparallel black hole.Torsion flux vanishes along the cosmic string itself.It is shown from Cartan equation in differential forms that the spins inside the black hole are polarized along the torsion string.Torsion string seems to be confined inside the black hole.
\end{abstract}
\vspace{0.5cm}
\begin{center}
\Large{PACS number(s):0450}
\end{center}
\newpage
\pagestyle{myheadings}
\markright{\underline{Torsion strings inside Static Black Holes in Teleparallel Gravity}}
\paragraph*{}
\section{Introduction}
Recently Pereira et al \cite{1} built a Kerr black hole \cite{2} in Einstein's teleparallel gravity \cite{3} and computed the Lense-Thirring effect in this case.It is also interesting to notice that the effects of tensorial and axial parts of torsion cancel in the equation of motion and there is no effect of torsion outside the static Schwarzschild black hole.In this letter we also consider teleparallel Schwarzschild solution but on a complete different context.First we use the method of Cartan's calculus of differential forms instead of the tetrad method used by Pereira et al. \cite{1}.In the second place we use a system of a cosmic torsion string ,not a spinning string \cite{4} passing through the center of a static black hole as given in Vilenkin and Shellard \cite{5}.The use of differential forms and the computation of torsion 2-forms allows us to compute the torsion flux integrals in the teleparallel case.Since it is possible to compute the relation between the torsion forms and the spin 3-forms is possible to show that the torsion string inside the black hole are polarized along the the cosmic string itself.A similar result has been obtained just for the system of cosmic strings and spin polarised matter around it by Garcia de Andrade \cite{6}.It is also interesting to point out that the problem of torsion singularities inside the black hole is also computed and we found that some of the components of torsion are singular on the Schwarzschild radius $r=2Gm$ while others components of torsion form are finite.Therefore to decide whether the singularity is on the Schwarzschild radius or at the center of the black hole like happens in the curvature case in Einstein's general relativity, one has to compute a torsion invariant analogous to the case of Kretschmann curvature scalar $R^{ijkl}R_{ijkl}=\frac{2m}{r^{3}}$ where $R_{ijkl}$ is the Riemann curvature tensor and ${i,j=0,1,2,3}$.It is shown from this computation that the true torsion singularity is at the center of blach hole which is also the locus of the torsion cosmic strings.In section $2$ we discuss the torsion singularities inside the Black Hole and compute the torsion components by using the teleparallel condition.In section $3$ the torsion flux is computed along with the spin polarisation direction of the spins inside the black hole.It is also interesting to note that this system maybe used in the case for example of neutrons stars with torsion strings inside to detect torsion since as it is well known this system possess about $10^{40}$ polarized spins \cite{7}.Just to give an idea of how big this figure is the ferromagnetic material which possess the highest spin density on Earth \cite{8} is above $10^{23}$ polarised spins! 
\section{Torsion singularities inside Blach Holes}  
Let us consider the spacetime metric describing the black hole with a cosmic string torsion inside 
\begin{equation}
ds^{2}=(1-\frac{2m}{r})dt^{2}-(1-\frac{2m}{r})^{-1}dr^{2}-r^{2}d{\Omega}^{2}
\label{1}
\end{equation}
where now the solid angle $d{\Omega}^{2}=d{\theta}^{2}+(1-8G{\mu})sin^{2}{\theta}d{\phi}^{2}$ where ${\mu}$ is the string mass and m is the Black hole mass.In terms of Cartan exterior differential forms the metric (\ref{1}) can be expressed as 
\begin{equation}
ds^{2}=({\omega}^{0})^{2}-({\omega}^{1})^{2}-({\omega}^{2})^{2}-({\omega}^{3})^{2}
\label{2}
\end{equation}
where the basis 1-forms ${\omega}^{i}$ are given by
\begin{equation}
{\omega}^{0}=(1-\frac{2m}{r})^{\frac{1}{2}}dt
\label{3}
\end{equation}
\begin{equation}
{\omega}^{1}=(1-\frac{2m}{r})^{-\frac{1}{2}}dr
\label{4}
\end{equation}
\begin{equation}
{\omega}^{2}=rd{\theta}
\label{5}
\end{equation}\begin{equation}
{\omega}^{0}=(1-8G{\mu})^{\frac{1}{2}}sin{\theta}d{\phi}
\label{6}
\end{equation}
From the Cartan structure equations
\begin{equation}
Q^{i}=d{\omega}^{i}+{\omega}^{i}_{j}{\wedge}{\omega}^{j} 
\label{7}
\end{equation}
where ${\omega}^{i}_{j}$ is the connection one-form and
\begin{equation} 
R^{i}_{j}=R^{i}_{jkm}({\Gamma}){\omega}^{k}{\wedge}{\omega}^{m}=d{\omega}^{i}_{j}+{\omega}^{i}_{k}{\wedge}{\omega}^{k}_{j}
\label{8}
\end{equation}
Here ${\wedge}$ is the exterior product of forms symbol and $R^{i}_{jkl}({\Gamma})$ are the components of the Riemann-Cartan geometry curvature tensor and $R^{i}_{j}$ is the curvature $2-form$.Rewriting the metric (\ref{1}) in the differential forms language one obtains
\begin{equation}
ds^{2}={\eta}_{ij}{\omega}^{i}{\omega}^{j}
\label{9}
\end{equation}
where ${\eta}_{ij}=diag(+1,-1,-1,-1)$ is the tetrad Minkowski metric.By making use of the teleparallel condition $R^{i}_{jkl}({\Gamma})=0$ into the equation (\ref{8}) we notice that the constraint 
\begin{equation}
{\omega}^{i}_{j}=0
\label{10}
\end{equation}
fulfills the teleparallel condition.Here we addopt this stronger teleparallel condition which also has been addopted by Letelier \cite{5} in the construction of torsion loops in teleparallel spacetimes.By using the condition (\ref{10}) into equation (\ref{7}) one obtains the torsion 2-form in the form
\begin{equation}
Q^{i}=d{\omega}^{i}=T^{i}_{jk}{\omega}^{j}{\wedge}{\omega}^{k}
\label{11}
\end{equation}
where $T^{i}_{jk}$ are the components of the Cartan's torsion tensor.Applying this simple expression to the expressions for the basis one-forms of the black hole torsion string system above one obtains after a quick computation one obtains the components of the torsion tensor as
\begin{equation}
T^{0}_{10}=\frac{\frac{Gm}{r^{2}}}{(1-\frac{2Gm}{r})^{\frac{1}{2}}} 
\label{12}
\end{equation}
\begin{equation}
T^{2}_{12}=\frac{1}{r}(1-\frac{2Gm}{r})
\label{13}
\end{equation}
\begin{equation}
T^{3}_{23}=\frac{cot{\theta}}{r} 
\label{14}  
\end{equation}
From these last expressions one notes that some components of torsion are singular at the Schwarzschild radius and others are singular at the center of the Black Hole.To decide 
which locus of singularity represents the true singular behaviour and which represents simply a coordinate singularity, we compute the torsion quadratic invariant $T_{ijk}T^{ijk}$ which is the analogous to the curvature scalar invariant in Riemannian geometry.In teleparallel gravity this is the most fundamental tensor invariant since there is  no way to build the Riemann-Cartan curvature invariant since this vanishes by the definition of teleparallelism.Therefore the scalar torsion invariant is
\begin{equation}
T^{ijk}T_{ijk}=-2[\frac{(\frac{Gm}{r^{2}})^{2}}{(1-\frac{2Gm}{r})}+\frac{(1-\frac{2Gm}{r})}{r^{2}}+(1-8G{\mu})\frac{ctg^{2}{\theta}}{r^{2}}]  
\label{15}  
\end{equation}
From this expression shows clearly that at the Schwarzschild radius the torsion invariant \cite{13} is $T^{ijk}T_{ijk}$ also vanishes as well as at the center of the black hole.Therefore there is no simple way to detect torsion singularities.The only thing we can say so far is that torsion is confined inside the Black Hole.Or Black Hole using Wheeler's therminology would have no torsion hair!On the next section we compute the torsion flux to have a better idea of the behaviour of torsion inside the Black Hole.
\section{Spin Polarised particles inside Black Holes from Torsion Strings}
By making use of the mathematical apparatus of the previous section we shall compute in this section the torsion flux according to Anandan \cite{9} as
\begin{equation}
\int_{\Sigma}{Q^{i}}=\int{{\omega}^{i}}
\label{16}
\end{equation}
where ${\Sigma}$ is the surface across which the torsion flux is computed and the second integral is the line integral due to the Stokes theorem.It is easy to note from the expressions for the torsion forms $Q^{i}$ that the only nonvanishing torsion flux is given by the integral
\begin{equation}
\int{\Sigma}{Q^{i}}=\int{{\omega}^{3}}=(1-8G{\mu})^{\frac{1}{2}}sin{\theta}\int{d{\phi}}=2{\pi}(1-8G{\mu})^{\frac{1}{2}}sin{\theta}
\label{17}
\end{equation}
Note that from this formula there is no torsion flux along the cosmic string ,where ${\theta}$ takes the value $0$ or $2{\pi}$.Therefore there is torsion flux along the direction around the 
cosmic torsion string as a torsion vortex in the case of superfluids.Note also that for a torsion string of mass ${\mu}=\frac{1}{8G}$ there is no torsion flux around this direction and no torsion flux at all!We could imagine this case as the system of black hole and torsion strings would shield totally the effects of torsion!Let us now make use of Cartan's equation relating the spin 3-form to the torsion 2-form to investigate the polarisation effect of the cosmic torsion string on the spinning particles of the spinning fluids of black holes.There is a small controversy if the teleparallel theories should possess spin density but here we addopt the point of view that the spin density is allowed in teleparallel gravity \cite{10}.From Cartan's equation
\begin{equation}
s_{ij}=\frac{-1}{8{\pi}G}[Q^{k}{\wedge}{\omega}^{l}{\epsilon}_{ijkl}]
\label{18}
\end{equation}
since the the only nonvanishing component of the torsion form is 
\begin{equation}
Q^{3}=\frac{cot{\theta}}{r}{\omega}^{2}{\wedge}{\omega}^{3}
\label{19}
\end{equation}
we can easily obtained the component of spin we wish by simply substituting formula (\ref{19}) into expression (\ref{18}) to yield
\begin{equation}
s_{12}=\frac{-1}{8{\pi}G}[\frac{cot{\theta}}{r}{\omega}^{2}{\wedge}{\omega}^{3}{\wedge}{\omega}^{0}{\epsilon}_{1230}]
\label{20}
\end{equation}
which shows that the polarisation of the spinning particles around the torsion string inside the black holes occur along the torsion string direction in analogy with the magnetic field which orients the magnetized particles along the magnetic field direction.Note that from expression (\ref{20}) the spin density seems to be singular at $r=0$.This nevertheless does not imply a Dirac delta distribution since off the torsion string there is matter inside the black hole in the form of spinning fluid in teleparallel gravity.
\section{Conclusions}
Torsion strings immersed on static black holes are shown to induce spin polarization of black hole spinning fluid along the torsion string direction itself.This is analogous to the polarisation of magnetic moment polarisation in magnetic materials in condensed matter physics.It is possible that in near future other types of torsion defects \cite{11} other than the torsion strings may be constructed inside teleparallel black holes.
\section*{Acknowledgments}
I am very much indebt to                                                   P.S.Letelier and J.G.Pereira for helpful discussions on the subject of this paper.Financial support from Universidade do Estado do Rio de Janeiro (UERJ) and CNPq. is grateful acknowledged.

\end{document}